\documentclass[prl,twocolumn,superscriptaddress,nofootinbib,amsmath,amssymb]{revtex4}
\usepackage{graphicx}
\usepackage{dcolumn}
\usepackage{lscape}
\usepackage{bm}
\usepackage{latexsym,epsfig}
\usepackage[usenames,dvipsnames]{color}
\usepackage{hyperref} 
\usepackage{verbatim}  
\usepackage{soul}   

\def\beq{\begin{equation}}
\def\eeq{\end{equation}}
\def\bea{\begin{eqnarray}}
\def\eea{\end{eqnarray}}

\def\eqref#1{Eq.~(\ref{eq:#1})}
\newcommand*{\eqlab}[1]{\label{eq:#1}}
\newcommand*{\figref}[1]{Fig.~\ref{fig:#1}}
\newcommand*{\figlab}[1]{\label{fig:#1}}

\def\VYP#1#2#3{{\bf #1}, #3 (#2)}  
\def\ApP#1#2#3{Astropart.~Phys.~\VYP{#1}{#2}{#3}}
\def\NIM#1#2#3{Nucl.~Inst.~Meth.~\VYP{A#1}{#2}{#3}}

\def\NPB#1#2#3{Nucl.~Phys.~B~\VYP{#1}{#2}{#3}}

\newcommand{\etal}{\mbox{\textit{et al.}}}                       %

\begin{document}


\title{Coherent Cherenkov Radiation from Cosmic-Ray-Induced Air Showers}

\author{K.D. de Vries}
\affiliation{Kernfysisch Versneller Instituut, University
of Groningen, 9747 AA, Groningen, The Netherlands}

\author{A.M. van den Berg}
\affiliation{Kernfysisch Versneller Instituut, University
of Groningen, 9747 AA, Groningen, The Netherlands}

\author{O. Scholten}
\affiliation{Kernfysisch Versneller Instituut, University
of Groningen, 9747 AA, Groningen, The Netherlands}

\author{K. Werner}
\affiliation{SUBATECH,
University of Nantes -- IN2P3/CNRS-- EMN,  Nantes, France}
\date{\today}

\begin{abstract}
Very energetic cosmic rays entering the atmosphere of the Earth will create a plasma cloud moving with almost the speed of light. The magnetic field of the Earth induces an electric current in this cloud which is responsible for the emission of coherent electromagnetic radiation. We propose to search for a new effect: due to the index of refraction of air this radiation is collimated in a Cherenkov cone. To express the difference from usual Cherenkov radiation, i.e.\ the emission from a fast moving electric charge, we call this magnetically-induced Cherenkov radiation. We indicate its signature and possible experimental verification.
\end{abstract}

\maketitle

\section{Introduction}

It is a well-known fact that an emitting source moving at a velocity exceeding the wave-propagation velocity in the medium will induce a shock-wave. Prime examples of this phenomenon are the sonic boom emitted by a super-sonic airplane, a bow-wave from a moving ship, and Cherenkov radiation emitted by an electric charge moving at almost the vacuum speed of light in a medium with an appreciable index of refraction. In this work we show that a similar effect occurs  in the emission from a fast moving electric current. It is suggested that this effect manifests itself in the emission of electromagnetic waves from particle cascades in the atmosphere of the Earth initiated by ultra-high energy (UHE) cosmic rays, with energies in excess of $10^{17}$\,eV.

An UHE cosmic ray entering the atmosphere of the Earth creates a cascade of particles, called an extensive air shower (EAS). In this cascade there are copious amounts ($>10^6$, depending on the initial energy of the cosmic ray) of electrons and positrons forming a small plasma cloud. This cloud, with a typical size of less than 1\,m, moves with almost the light velocity towards Earth.
The magnetic field of the Earth, by means of the Lorentz force, induces a drift velocity for the leptons which is perpendicular to the direction of the initial cosmic ray and opposite for electrons and positrons. As a result an electric current is created in the fast moving plasma cloud. The strength of the induced electric current is roughly proportional to the number of charged particles in the plasma cloud as the induced drift velocity is varying little with height. This macroscopic picture~\cite{Sch08} was recently confirmed~\cite{Hue10-Arena} to agree with a microscopic description~\cite{Lud10-Arena}.
Even when the index of refraction of air would be equal to that of vacuum this varying electric current emits electromagnetic waves and coherent emission occurs at a wavelength longer than the size of the charge cloud, i.e.\ for radio frequencies $\nu< 300$\,MHz~\cite{Sch09-Arena}. The geomagnetic emission mechanism~\cite{Kah66} has been confirmed from data~\cite{Fal05}. In addition to the induced current the plasma cloud has a net charge excess which also radiates. The polarization direction of radiation distinguishes the emission due to the charge excess and the geomagnetic current~\cite{dVries10,Schoorl11}. Since charge-excess radiation is generally smaller in intensity we will concentrate in this work on geomagnetic emission.

As is well known the propagation speed of electromagnetic waves is $c/n$ where $c$ denotes the velocity of light in vacuum. In this work we investigate the effect of the index of refraction of air, $n$, on the emission following Ref.~\cite{Wer08}. The effects of Cherenkov radiation from EAS have also been addressed in Ref.~\cite{Kalmykov2006347}. In this work we show that for realistic values for $n$ the Cherenkov effect introduces distinct features in the ground pattern of the emitted radiation.

\section{Radio wave emission}

As described, a cosmic ray entering the atmosphere induces an EAS, a plasma cloud moving at almost the light velocity, where the magnetic field of the Earth induces a net electric current in the plasma~\cite{Sch08}. From classical electrodynamics~\cite{Jac-CE} we obtain the Li\'enard-Wiechert potentials for a point source following a trajectory $\vec{\xi}(t')$ and an observer at rest at $(t,\vec{x})$,
\beq
A^{\mu}_{PL}(t,\vec{x}-\vec{\xi}(t'))=
\frac{\mu _0}{4\pi} \frac{J^\mu_{PL}}{|{\cal D}|}\Bigr{|}_{t=t'} \;,
\eqlab{vec-pl}
\eeq
where $t'$ denotes the retarded time corresponding to the time the signal was emitted from the moving charge distribution and ${\cal D}$ is the retarded distance~\cite{Wer08}. In the point-like approximation (denoted by $_{PL}$) where the size of the plasma cloud is infinitesimal the four-current is,
\beq
J^\mu_{PL}(t',\vec{x})=J^\mu(t')\delta^{3}(\vec{x}-\vec{\xi}(t')) \;,
\eeq
where $\vec{\xi}(t')=-ct' \vec{e}_z$ and $J_{PL}$ carries a longitudinal component due to the net charge excess (which we will ignore) and a transverse component due to the {opposite drift of electrons and positrons induced by the Earth magnetic field. The transverse component is} proportional to the number of charged particles at a given height, $J^\perp(t') \propto N_e(z)$ where $z=-c\,t'$ is the distance to the Earth's surface as measured along the shower path.

The electromagnetic fields at the observer are given as,
\bea
E^{i}&=&c(\partial^{i}A^{0}-\partial^{0} A^{i}) \;,
\eqlab{e-field}
\eea
{with $i=x,y,z$, an observer located} at $\vec{x}=(x,y,z=0)$, and where at $t=t'=0$ the shower hits the Earth. The distance $d=\sqrt{x^2+y^2}$ is equal to the transverse distance to the shower axis. The velocity of the charge cloud is $c\beta$ where we set $\beta=1$ in the following. For the special case $n=1$, the retarded distance ${\cal D}=\sqrt{(-\beta c t)^2+(1-n^2\beta^2)d^2}$ reduces to ${\cal D}=\beta c t$ which is finite.

The retarded time is defined by the light-cone condition
$
c(t-t')=L(\vec{x} , \vec{\xi}(t')) \;,
$
where the distance $L=L(\vec{x},\vec{\xi})$ is the optical path length between the source located at  $\vec{\xi}$ and the observer at $\vec{x}$. In reality the index of refraction of air depends on density and thus height, $n=n(z)$, hence light will follow curved trajectories where $L$ is the integral $\int ds$ along the light curve from $\vec{\xi}$ to $\vec{x}$.

\begin{figure}[!htb]
\centerline{ \includegraphics[width=.4\textwidth, keepaspectratio]{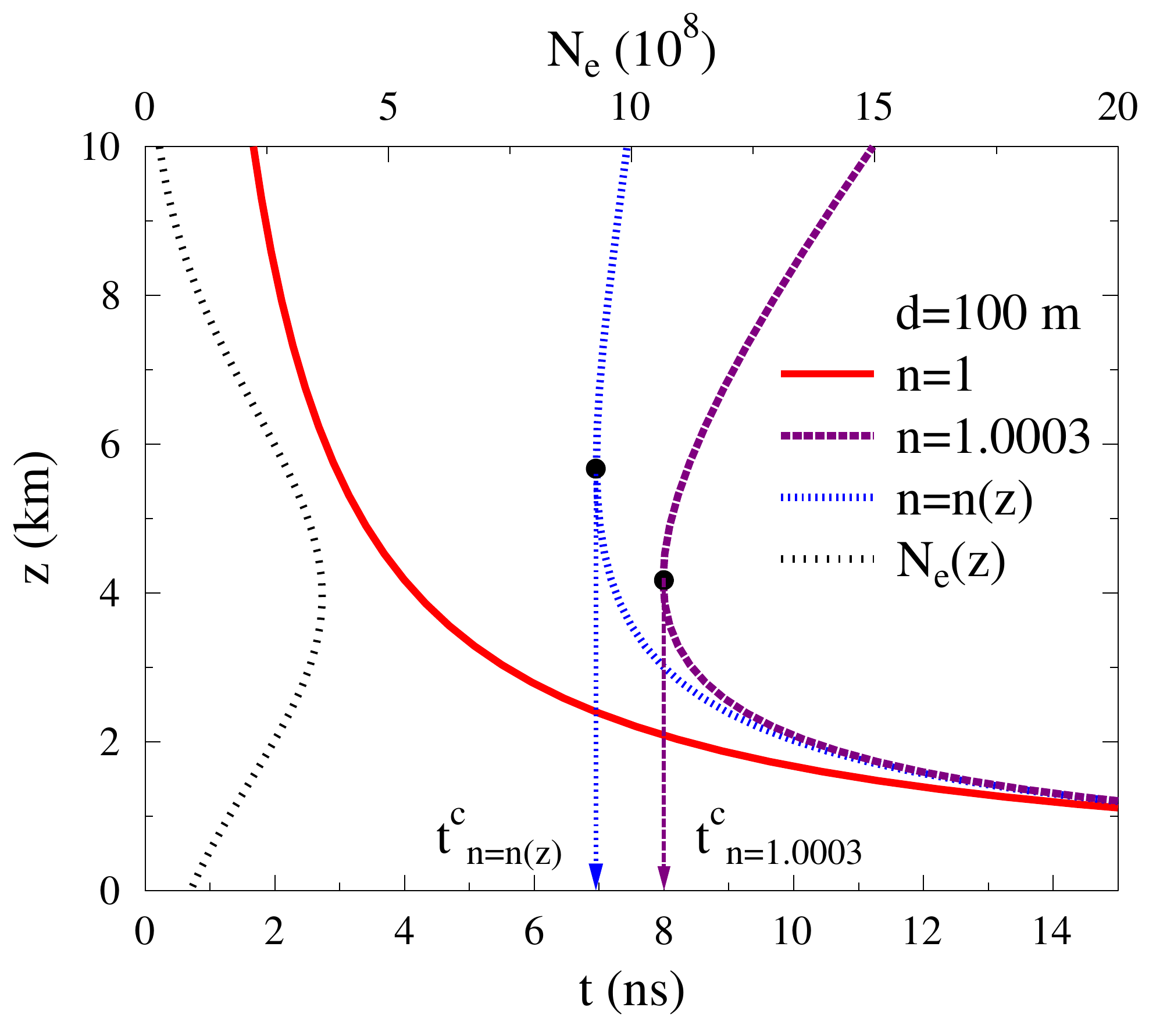}}
\caption{The  emission height $z=-c\,t'$ as function of the observer time $t$ for three different values of the index of refraction. The dashed line gives the shower-profile as function of $z$ for a $E=5\times 10^{17}$\,eV proton-induced shower. }
\figlab{trett}
\end{figure}

In \figref{trett} the emission height, $z=-c\,t'$, is plotted as a function of the observer time $t$ for an observer 100\,m from the shower axis for three choices for the index of refraction $n$. For the case $n=1$ the plot of the retarded time (red drawn curve) shows that the retarded time is a single valued function and that the earliest signals come from the top of the shower.
For an index of refraction deviating from unity ($n=1.0003$, $n=n(z)$) the function is composed out of several branches (dashed-magenta and dotted-blue curves in \figref{trett}) limited by the critical times $t=t_c$, where the derivative $dt'/dt$ becomes infinite. In~\cite{Wer08} it was already shown that a singularity in this derivative is related to a singularity in our vector potential since ${\cal D}=R/(dt'/dt)$. This singularity (the branch point) is well known and corresponds to Cherenkov emission. Since the critical time is the time the Cherenkov radiation is seen by the observer, it is henceforth called the Cherenkov time.
In the case of a constant and finite refractivity $N=n-1$ the Cherenkov time $t_c$ is given as,
\beq
c t_c=d\,\sqrt{(n^2-1)} \;,
\eqlab{tk}
\eeq
with corresponding retarded time $-t'_c=t_c/(n^2-1)$. The retarded time can readily be converted into an emission height $z_c=-c\, t'_c$ and it can be seen that the expressions are consistent with the expression of the angle for Cherenkov emission, $\cos \theta_c=1/n$. For the general case it has to be calculated numerically.

To calculate the realistic pulse form for $n\neq 1$ it is essential to include the fact that the emitting plasma cloud has a finite size,
\beq
A^{\mu}_w(t,\vec{x})   = \! \int \! d^{2}\vec{r} \! \int \! dh \,
\rho(h,\vec{r}) \, A^{\mu}_{PL}(t,\vec{x}-\vec{\xi})
\eqlab{vec-pot}
\eeq
where $\vec{r}=r_1\vec{e}_x + r_2\vec{e}_y$ is the relative transverse coordinate to the shower axis and the source is at the position $\vec{\xi}(t',h,r_i) = (-c\,t'+h)\,\vec{e}_z + \vec{r}$.
The density profile is parameterized as~\cite{Agn97,Hue03}
\beq
w(r,h)=2\pi r \rho(h,\vec{r})= {\cal N} (r+r_0)^{-3.5} h\, e^{-2h/L}
\eeq
at a fixed shower time $t'$ 
and where $h$ is the longitudinal distance from the shower front that, by definition, moves with the vacuum speed of light $c$. The normalization constant, ${\cal N}$, is chosen such that $\int \! dr \! \int \! dh \, w(r,h)=1$. Positive values of $h$ mean a position behind the shower front, hence $w$ is zero for values of $h<0$.
The parameters are taken as $r_0=80$\,m, $L=0.5$\,m following the results of simulations using the cascade mode of the  CONEX-MC-GEO shower simulation and analysis package~\cite{Wer08}.
The results of these simulations indicate that the pancake thickness parameter $L$ has to be considerably smaller than the value of $L=3.9$\,m used in previous calculations~\cite{dVries10}.
Retaining the terms pertinent to geomagnetic radiation and using partial integration to have the derivatives in \eqref{vec-pot} operate on the density distribution exclusively, the electromagnetic field can be expressed as
\beq
\eqlab{e-field-full}
E^{i} = \! - \frac{\mu_0 c}{4\pi}\int \! d^2\vec{r} \! \int \! dh \frac{1}{|{\cal D}|} \left(
	\frac{\partial w}{\partial h} J^{i}_{PL}
	+ w \frac{\partial {J}^{i}_{PL}}{\partial t'} \right)\;. 
\eeq
The complication in \eqref{e-field-full} is now reduced to the integration of inverse square-root divergencies over smoothly varying functions giving finite results. We will argue later that the first term is important for Cherenkov radiation while the second is dominant for $n=1$.

\begin{figure}[!htb]
\centerline{
\includegraphics[width=.30\textwidth, keepaspectratio]{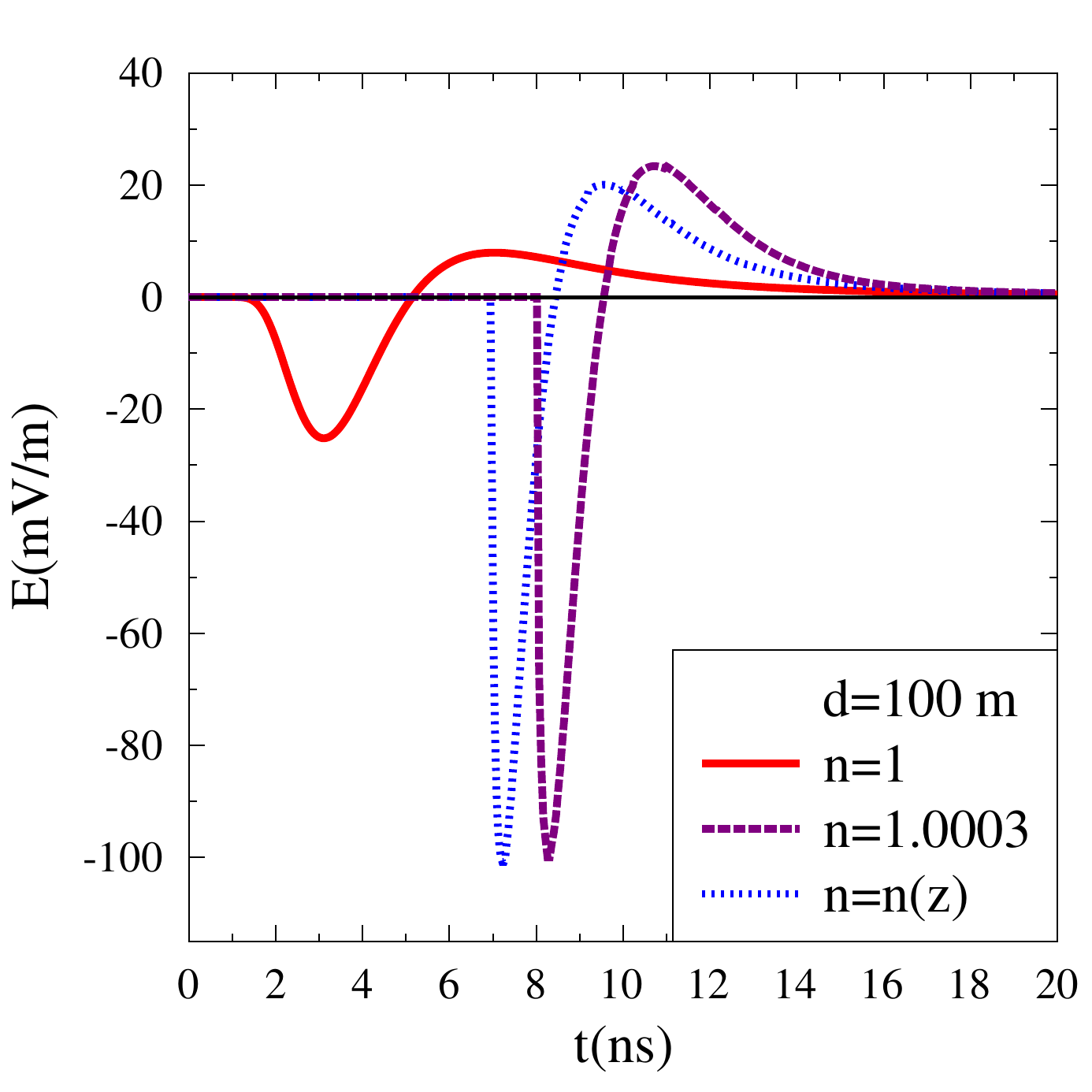}}
\caption{Pulse due to geomagnetic radiation at $d=100$\,m from the core for different values of the index of refraction as discussed in the text, $n=1$, $n=1.0003$ fixed, and $n=n(z)$ realistic. In this calculation the lateral extent of the shower is ignored.}
\figlab{gm-r0-d100}
\end{figure}

To get a better appreciation of the structure of the pulse and the influence of the index of refraction we will discuss in some detail the results for an observer at a distance $d=100$\,m from a shower with $E=5\times 10^{17}$\,eV at an angle $\theta=27^\circ$. To simplify the discussion we will -for the time being- ignore the radial extent of the charge cloud. 
Essential for the structure of the pulse is the longitudinal shower profile $N_e(z)$, as shown by the thinly-dotted black curve in \figref{trett}.

For the case that $n=1$ the denominator ${\cal D}$ is a smoothly changing function over the integration regime and the dominant contribution is derived from the last term in \eqref{e-field-full}. Since the derivative of the shower profile reaches a maximum at $z=6$\,km it can be seen from \figref{trett} that the peak of the pulse occurs at $t \approx 3$\,ns which agrees with the result in \figref{gm-r0-d100}. The integral of the first term, being a derivative, almost vanishes.

For the cases that $n\neq 1$ the reasoning is rather different. The denominator ${\cal D}$ vanishes at the Cherenkov time, indicated by the vertical arrows in \figref{trett}, and thus $1/{\cal D}$ varies strongly as function of $h$. The contribution from the first term in the integral \eqref{e-field-full} is large (in contrast to the case $n=1$) and results in an enhanced contribution from the corresponding emission height. The pulse height will thus be proportional to $N_e(z_c)$ while the peak is observed a little after $t_c$ in agreement with the results shown in \figref{gm-r0-d100}.  Since $J^\perp(t'_c) \propto N_e(z_c)$ is large for $d=100$\,m this results in a strong pulse. The remaining contributions in the integral are of secondary importance in this case.

In reality the refractivity is equal to $N(0)=3\times 10^{-4}$ at ground level and decreases exponentially with height, $N(z)=n(z)-1$. The values of the retarded time thus lie in between those obtained with $n=1$ and $n=1.0003$ and are shown as the dotted-blue curve in \figref{trett}.
Also for this case there is a clear Cherenkov time which is slightly smaller than that for $n=1.0003$ 
resulting is a very similar pulse as for the case of a fixed-finite $n$. 

The main effect of including the radial extent of the shower is to smooth the time structure of the pulse and thus to wash-out some of the effects of $n\neq 1$. The third panel of \figref{gm-r-dall} gives the complete geomagnetic result and should be compared with \figref{gm-r0-d100}. The differences between the three different choices for the index of refraction have diminished, instead of being three times as large the pulse height is increased by a factor two only.

\begin{figure}[!htb]
\centerline{	\includegraphics[width=.25\textwidth, keepaspectratio]{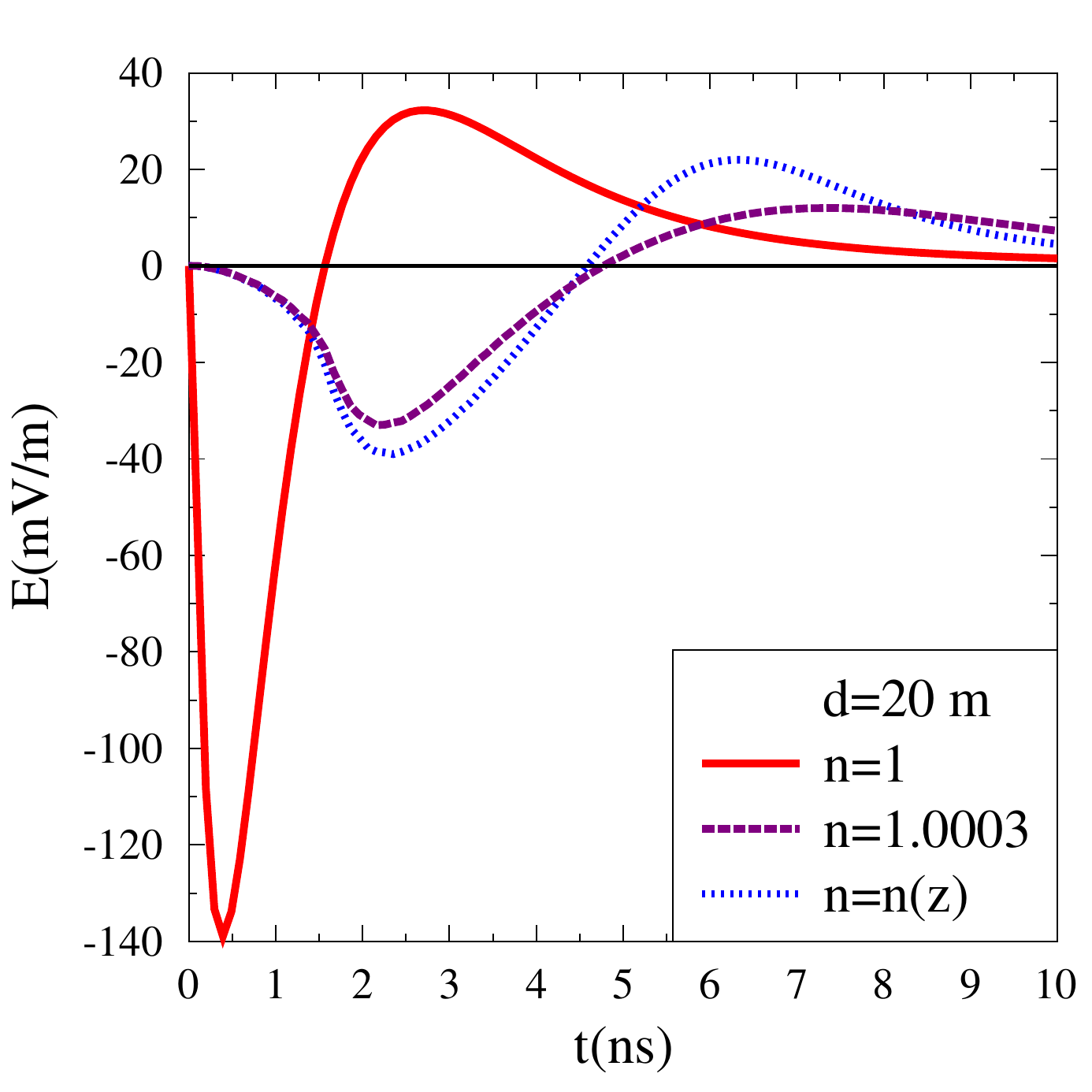}%
			\includegraphics[width=.25\textwidth, keepaspectratio]{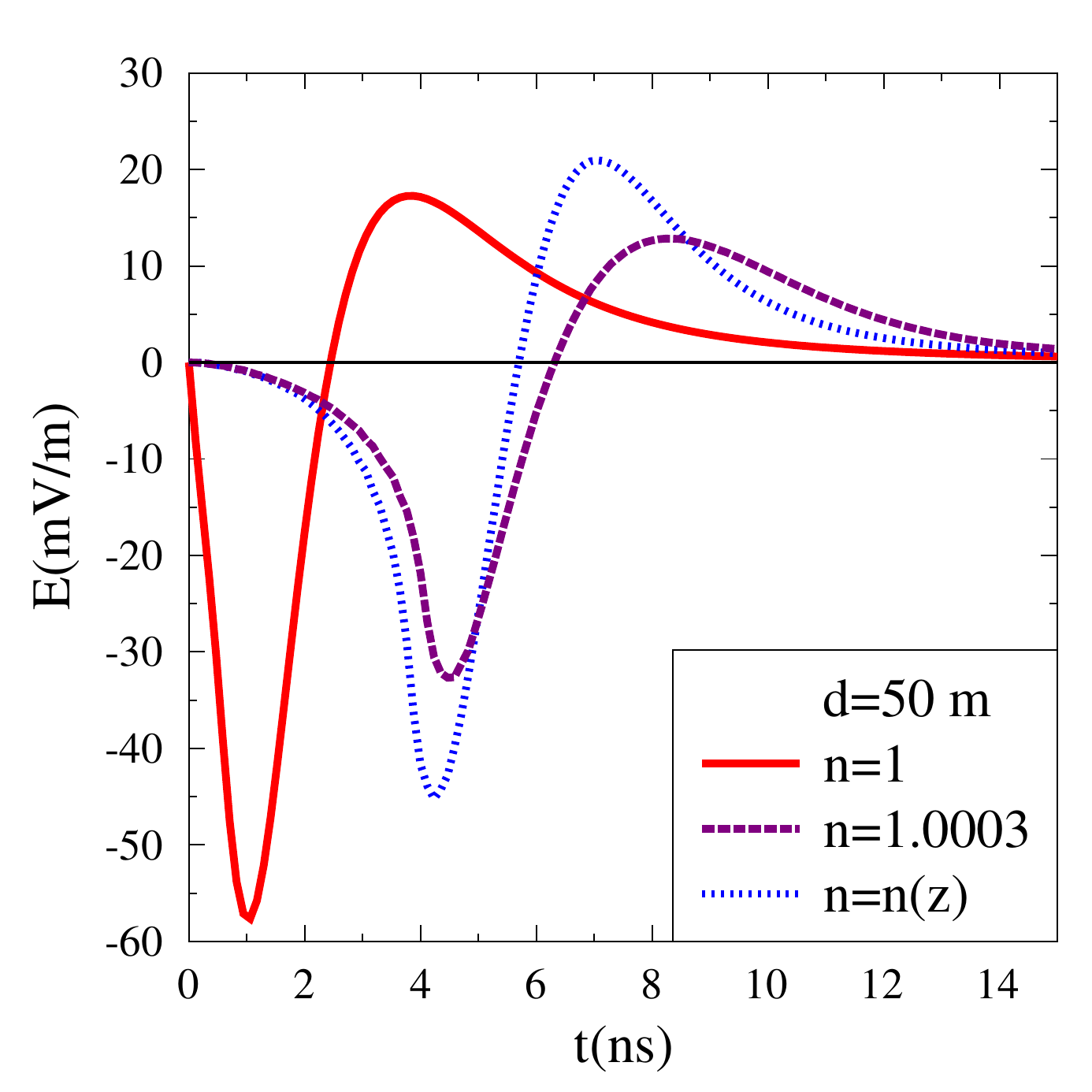}	}
\centerline{	\includegraphics[width=.25\textwidth, keepaspectratio]{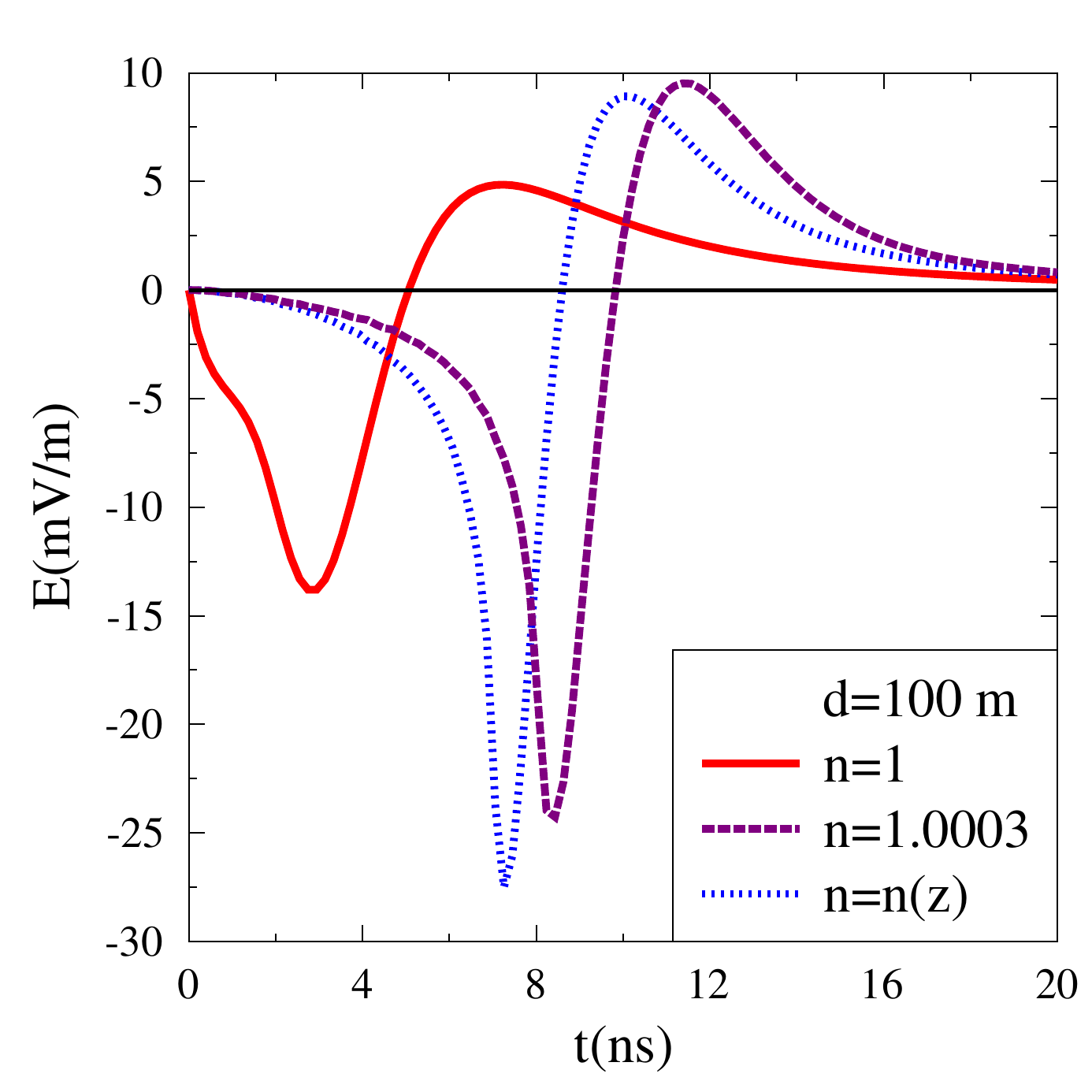}%
			\includegraphics[width=.25\textwidth, keepaspectratio]{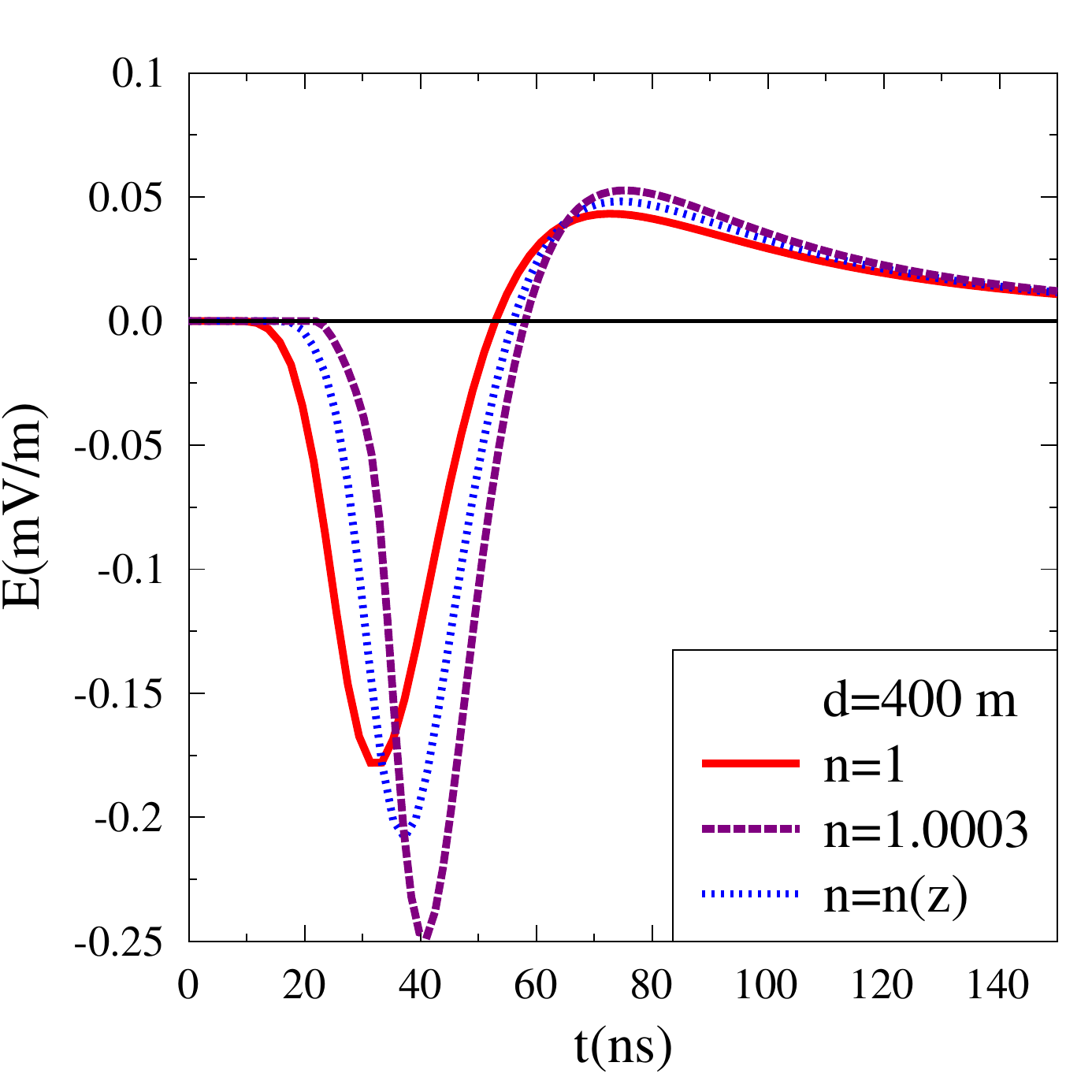}}
\caption{Pulse due to geomagnetic radiation at at several distances from the core for different values of the index of refraction as discussed in the text ($n=1$, $n=1.0003$ fixed, and $n=n(z)$ realistic), including the lateral extent of the shower.}
\figlab{gm-r-dall}
\end{figure}

The Cherenkov effect is strongly distance dependent as shown in \figref{gm-r-dall}. As expressed in the discussion following \eqref{tk} emission at the Cherenkov angle $\cos{\theta_c}=1/n$ relates a distance $d$ from the shower core to a particular emission height.
For an observer close to the shower, $d=20$\,m, this Cherenkov height lies well below the shower maximum where $N_e(z)$ has fallen considerably and the intensity of the emitted radiation is low. It increases with increasing distance to reach a maximum around $d=100$\,m where the Cherenkov point lies at the height of the shower maximum. At even larger distances the Cherenkov point lies above the shower maximum and the amplitude decreases rapidly. The position of the Cherenkov maximum thus reflects the shower maximum and is thus sensitive for the composition of the cosmic ray. For  $n=1$ the main strength of the pulse is emitted at the height where the derivative of the shower profile is large. Since the retarded distance, $|{\cal D}|$, corresponding to this height in the shower evolution increases for observers further away from the shower core, the amplitude of the pulse is a monotonically decreasing function with distance.

Calculations confirm that for charge-excess radiation (not reported here) the effect of the index of refraction is very similar as for geomagnetic radiation where there are subtle differences due to the somewhat different weighting over shower height. For the chosen geometry the geomagnetic effect is however dominant. This shows that the Cherenkov effect applies equally to radiation from a moving charge, to which it is usually applied, as to that from a moving electric current, which is at the focus here.

\section{Experimental verification}

The polarization of the radio signal, which distinguishes geomagnetic and charge-excess radiation, is not affected by the Cherenkov effect.
The principal signature of the Cherenkov effect is that the pulse at a certain distance, 100\,m for the present example, is considerably larger than it would have been for the case that $n=1$. This feature is especially clear from \figref{gm-r-dPH} where the height of the pulse 
is plotted at various distances from the core. 
At short distances the pulse height diverges for the case of $n=1$ where it should be noted that this divergence is strongly dependent on the pancake thickness parameter $L$, smaller values give a stronger divergence at $d=0$. Apart from an overall scaling factor the picture is not affected by $L$ for $d\gtrsim 50$\,m. For the realistic case where $n\neq 1$ a marked deviation is predicted with a clear enhancement in peak intensity at distances ranging from 50 till about 100\,m from the core. This peak results from the fact that the Cherenkov point lies close to the shower maximum for $d=$100\,m.

\begin{figure}[!htb]
\centerline{ \includegraphics[width=.35\textwidth, keepaspectratio]{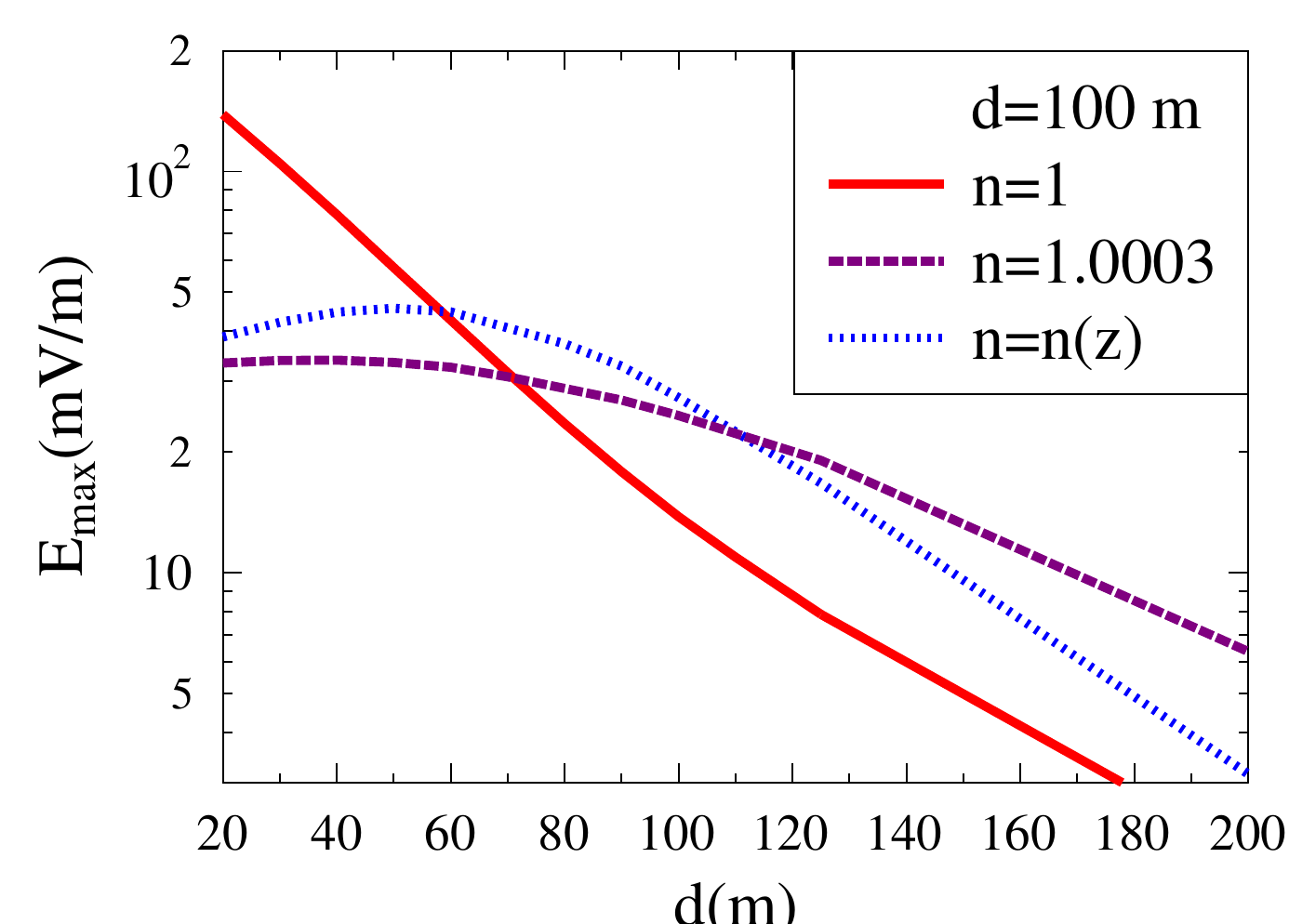} }
\caption{Pulse height as function of distance to the shower core for the three choices for the index of refraction as discussed in the text ($n=1$, $n=1.0003$ fixed, and $n=n(z)$ realistic).}
\figlab{gm-r-dPH}
\end{figure}

The feature shown in \figref{gm-r-dPH}, an increase of the amplitude of the geomagnetic radiation with distance for $n\neq1$ instead of the monotonic decrease predicted for $n=1$, remains for showers at a large angle with the vertical. For these inclined showers the shower maximum will occur at a larger values of $z$, the distance to the point of impact on Earth, and the maximum in the intensity will thus be observed at larger distances from the shower core, consistent with the angle for Cherenkov emission.

Some hints of this effect may have been seen in recent results from the LOPES~\cite{LOPES-flat} collaboration showing that for certain events the pulse height follows the trend shown by the dotted-blue line in \figref{gm-r-dPH}. More detailed measurements are necessary where full attention is given to polarization observables which are crucial to distinguish geomagnetic and charge-excess radiation. Such measurements are planned for new and renovated set-ups at LOPES, CODALEMA~\cite{CODALEMA}, and recently also new set-ups at the Pierre Auger Observatory (MAXIMA~\cite{Cop09}, AERA~\cite{AERA}), and LOFAR~\cite{LOFAR}.

\section{Acknowledgments}
This work is part of the research program of the 'Stichting voor Fundamenteel
Onderzoek der Materie (FOM)', which is financially supported by the 'Nederlandse
Organisatie voor Wetenschappelijk Onderzoek (NWO)'.

\end{document}